\documentclass[12pt]{article}
\usepackage{graphicx}
\begin{document}
\begin{titlepage}
\title{Energy of Gravitational Radiation and the Background Energy
of the Space-Time}

\author{J. W. Maluf$\,^{(a)(1)}$, F. L. Carneiro$\,^{(b)(2)}$, \\
\bigskip
S. C. Ulhoa$\,^{(c)(1)}$ and J. F. da Rocha-Neto$\,^{(d)(1)}$ \\
{(1)}\, Universidade de Bras\'{\i}lia, Instituto de F\'{\i}sica, \\
70.910-900 Bras\'{\i}lia DF, Brazil\\
(2)\, Universidade Federal do Norte do Tocantins, \\ 
Centro de Ci\^encias Integradas, 77824-838 Aragua\'ina
TO, Brazil  }
\date{}
\maketitle
\begin{abstract}
We address the issue of gravitational radiation in the context of the
Bondi-Sachs space-time, and consider the expression for the gravitational
energy of the radiation obtained in the realm of the teleparallel
equivalent of general relativity (TEGR). This expression is independent
of the radial distance (i.e., of powers of $1/r$) and depends exclusively
on the functions $c(u,\theta,\phi)$ and $d(u,\theta,\phi)$, 
which yield the news functions ($u$ is the retarded time, $u=t-r$).
We investigate the mathematical and physical features of this energy 
expression in the simpler framework of axial symmetry. Once a burst of 
gravitational radiation takes place in a self gravitating system, that 
leads to a loss of the Bondi mass, gravitational 
radiation is emitted throughout the whole space-time. The existence 
and presence of this radiation in the background structure of the 
space-time is consistent with the analysis developed by 
Papapetrou, and Hallidy and Janis, who found no proof that
a gravitational system that emits a burst of gravitational radiation
is preceded and followed by two stationary gravitational field 
configurations, namely, it seems
that it is impossible for a gravitational system, which is initially 
stationary, to return to a stationary state after emitting a burst of 
axially symmetric gravitational radiation, in which case the space-time
is not even asymptotically stationary. Therefore, it is plausible that
the gravitational energy of radiation is present in the background
structure of the space-time, and this is the energy predicted in the TEGR.
This analysis lead us to conjecture that the noise detected in the large 
terrestrial gravitational wave observatories is intrinsically related to
the background gravitational radiation.
\end{abstract}
\thispagestyle{empty}
\vskip 1.0cm
\thispagestyle{empty}

\noindent (a) jwmaluf@gmail.com, wadih@unb.br\par
\noindent (b) fernandolessa45@gmail.com\par
\noindent (c) sc.ulhoa@gmail.com\par
\noindent (d) jfrocha@unb.br, jfrn74@yahoo.com \par

\end{titlepage}
\tableofcontents

\section{Introduction}
Gravitational radiation is a phenomenon that occurs due to a loss of mass of
gravitational systems, and is a consequence of violent astrophysical 
processes that result in gravitational field configurations very different
from the initial configuration. The loss of mass leads to a change in the
space-time geometry, and this change may generate ripples
in the space-time in regions far from the source, as well as a drastic
modification of the configuration of the very source.
In the context of the Bondi-Sachs formalism
\cite{Bondi,Sachs}, which we adopt here, the asymptotic form of the field 
variables (i.e., of the metric tensor components) is typically of powers of 
$1/r^n$, $n=1,2,3...$, where $r$ is a suitable radial coordinate. But the 
space-time is not strictly asymptotically flat because the components of the 
metric tensor depend on the retarded time $u=t-r$.

Let us suppose that the space-time of a certain astrophysical configuration 
is initially stationary, and suddenly it emits a pulse of gravitational 
radiation. The question about the existence of a final stationary 
configuration after the pulse of gravitational radiation, where the radiation
process of the source is finished, was investigated by 
Papapetrou \cite{Papapetrou}, and Hallidy and Janis \cite{Janis}, and 
independently verified by Hobill \cite{Hobill}. The results are not totally
 conclusive, but a proof of the existence of a final stationary
field configuration has not been reached. The work by Hallidy and Janis
\cite{Janis} showed that it is unlikely for a gravitational system to return 
to a stationary state after emitting a burst of axially symmetric 
gravitational 
radiation of finite multipole expansion.  The question of whether or not it 
can return to a stationary state is still open, and would perhaps require to 
abandon the case of axial symmetry, and include $\phi$-dependent field 
variables, as discussed in Ref. \cite{Janis}.
An alternative possibility, as they argue, consists in considering
an infinite multipole expansion of the function $c(u,\theta)$, which yields 
one of the news function and which will be defined ahead. In 
Refs. \cite{Papapetrou,Janis,Hobill}, the function $c(u,\theta)$ is 
precisely given by an infinite 
multipole expansion, and the authors of these references consider
only a finite expansion in their analyses. 
In contrast, in this article, we consider
the full multipole expansion suggested by these authors, and carry out a 
numerical analysis by stipulating some trial forms for $c(u,\theta)$, as
initial data for the configuration.

One can argue, as in Ref. \cite{Hobill}, that the Bondi formalism is unable 
to establish the existence of radiative transitions between two exactly 
stationary states. However, a mathematical formalism is not by itself 
responsible for the existence or not of radiation. It seems plausible to 
assert the existence of gravitational radiation in the background structure
of the space-time provided, to a certain extent, that this radiation is 
non-dissipative, and that it can be measured with the sophisticated 
instruments presently used to detect gravitational waves. Nowadays, it is 
agreed that in the whole history of the universe, uncountable astrophysical 
processes have taken place that generated an unspecified form of 
gravitational radiation. 

In this article, we argue that one possible manifestation of a Bondi-type
gravitational
radiation discussed above is given as a substantial contribution to the noise
detected in the large terrestrial gravitational wave laboratories such as 
LIGO, Virgo and KAGRA. These detectors are precisely designed to measure 
linearised gravitational waves, which is a form of gravitational field.
Thus, the detection of noise as a form of gravitational field cannot be ruled
out. In the recent literature about linearised gravitational waves, noise is
simply characterised (stipulated)
as due to non-astrophysical processes, such as 
instrumental glitches, environmental disturbances and mechanical resonances.
A good and recent discussion on this issue is provided in Ref. \cite{Gupta}
(and references therein).
The fact is that the exact nature of the noise is unknown. Therefore, the
possibility that the noise is also due to different forms of astrophysical
processes cannot be discarded. In Ref. \cite{Gupta}, the discussion about 
``missing physics" is made in order to better explain the measurement of
linearised gravitational waves, and the related noise in the observations.
As we will show, the expression of the 
gravitational energy of radiation obtained in the teleparallel equivalent of
general relativity (TEGR) clearly supports the conclusion discussed above,
regarding the relationship between noise and gravitational radiation. The
existence of such a form of gravitational energy permeated (and travelling) 
all over the universe is in agreement with the conclusions obtained by
Papapetrou \cite{Papapetrou}, and Hallidy and Janis \cite{Janis}, according
to which a stationary state, as a final gravitational field configuration, is 
likely not to exist.

The TEGR is a consistent formulation of the relativistic theory of 
gravitation, both from the Lagrangian and Hamiltonian points of view, as 
shown in two review articles \cite{TEGR-1,TEGR-2}. It is constructed out of a
set of tetrad fields $e^a\,_\mu(x)$, where $a$ and $\mu$ are Lorentz
(tangent space) and space-time indices, respectively, which acquire values
$a = (0), (1), (2), (3)$ and $\mu = 0, 1, 2, 3$. The field equations 
satisfied by the tetrad fields are equivalent to Einstein's field equations,
and are invariant under local Lorentz transformations.
The tetrad fields suffice to describe all mathematical and physical features
of the theory. The internal structure of the TEGR is suited to address the
definitions of energy-momentum and 4-angular momentum of the gravitational
field.
We have discussed several times in the literature on the 
fact that a flat spin connection is dispensable in the theory, and an
exhaustive discussion on this issue has been presented in Ref. \cite{Maluf1}.
A detailed discussion is also given in Ref. \cite{TEGR-2}.
We consider that the repetition of this discussion is unnecessary in
this article. Given that a full, complete presentation of the TEGR can be 
found in the review articles \cite{TEGR-1,TEGR-2}, with emphasis to the
second recent review, here we will just display a very short presentation of
the theory. For this reason, in Section 2 of the article we briefly 
present the TEGR and the definition of the energy-momentum of the 
gravitational field.

The crucial starting point for the present analysis are the results obtained
in Ref. \cite{Maluf2}. We have carefully applied the definition of the 
gravitational energy-momentum to the complete Bondi-Sachs space-time. The
total gravitational energy of the space-time is given by the integration of
the mass aspect $M(u,\theta,\phi)$, as usual, and by an extra expression not
previously considered in the literature. This extra expression is constructed
out of the functions 
$c(u,\theta,\phi)$ and $d(u,\theta,\phi)$, which yield the 
news functions and which, together with the mass aspect, constitute the three
independent functions that characterise the space-time and must be fixed as
initial data of the configuration. All metric components are constructed in
terms of these three functions. We recall that long time ago, the news 
functions were considered the radiating degree of freedom of the 
gravitational field. Nowadays, this interpretation has lost relevance, but 
the role of the news functions remains unexplored.

We will restrict considerations to an axially symmetric space-time, for
simplicity, and
obtain graphical expressions of the gravitational energy of the
radiation for several forms of the function $c(u,\theta)$. We will
show that all of them represent a pulse of gravitational energy propagating 
in space-time, that 
exhibit the same characteristic feature: they are oscillations 
of the total gravitational energy, possibly tiny oscillations,  with equal
positive and negative domains of gravitational energy along the propagating
axis determined by the retarded time $u$. The energy of the gravitational
radiation does not change the total value
of the Bondi energy $m(u)$ of the space-time, 
but leads to tiny propagating oscillations of the background gravitational 
energy, even in regions far from the source of the astrophysical events that
generated the radiation, namely, even in hypothetically, locally flat 
space-time regions. Thus, these tiny oscillations are everywhere present in
the universe, and could already have been detected by the large terrestrial
gravitational wave observatories in the form of noise. 

In Section 2 of this article, we also recall the main definitions of the
Bondi-Sachs space-time, as well as the expression of the gravitational energy
of radiation. This issue has been thoroughly presented in Section 6 of Ref.
\cite{TEGR-2}. Again, we will display a brief description of the subject, 
since all details may be found in Refs. \cite{TEGR-2,Maluf2} and, as long as
possible, we will dispense with repetitions. In Section 3 we will present our
main results.

\section{The TEGR and the Bondi-Sachs framework}

\subsection{A brief review of the TEGR}

The core of a theory are the field equations, from which the physical 
predictions are obtained. The TEGR is a theory that encompasses the features
of Einstein's general relativity with the idea of distant parallelism, or
teleparallelism, established by the Weitzenb\"{o}ck connection 
$\Gamma^\lambda_{\mu\nu}=e^{a\lambda}\partial_\mu e_{a\nu}$. If the covariant
derivative $\nabla_\mu V^\lambda$ of a certain vector field
$V^\lambda$ vanishes 
all over the space-time, where the covariant derivative is constructed out of
the Weitzenb\"{o}ck connection, then we say that the vector field is 
everywhere parallel in space-time, i.e., 
$V^a(x)=e^a\,_\mu(x)V^\mu (x)= V^a(x+dx)=e^a\,_\mu(x+dx)V^\mu (x+dx)$. 
The covariant derivative of the tetrad fields vanishes identically,
$\nabla_\mu e^a\,_\lambda$=0.  In this case, we may conclude that the whole
space-time has a structure similar to a crystalline lattice.

The curvature tensor constructed out of the Weitzenb\"{o}ck connection
vanishes, but not the torsion tensor,

\begin{equation}
\label{1}
T^\lambda_{\mu\nu}=e^{a\lambda}(\partial_\mu e_{a\nu}-
\partial_\nu e_{a\mu})\equiv e^{a\lambda}T_{a\mu\nu}\,.
\end{equation}
In our notation, the flat, tangent space metric tensor 
$\eta_{ab}=(-1, +1, +1, +1)$ is related to the contravariant components
of the space-time metric tensor $g^{\mu\nu}$ according to
$\eta_{ab}=e_{a\mu}e_{b\nu}g^{\mu\nu}$, as usual.

In view of Eq. (\ref{1}), we naturally define the space-time Burgers
vector $b^a$ according to 

\begin{equation}
\label{2} 
b^a={1\over 2} \int_S T^a\,_{\mu\nu} dx^\mu \wedge dx^\nu\
= \oint_C e^a\,_\mu dx^\mu\,,
\end{equation}
where $C$ is the contour of the open surface $S$.
We may envisage the space-time endowed with a density of defects 
(dislocation type defects), given by Eq. (\ref{2}), that expresses the 
deviation of the actual space-time from a flat space-time devoid of 
deformations. This alternative point of view of the 
space-time may be useful and interesting when considering certain
gravitational field configurations such as pp-waves or 
global space-time defects. However, the ultimate structure of the 
dynamics of the theory is really equivalent to Einstein's general 
relativity.

The Lagrangian density of the TEGR is precisely constructed out of the 
torsion tensor $T_{a\mu\nu}$. In vacuum, the well known expression is 
given by \cite{TEGR-1,TEGR-2}

\begin{eqnarray}
L(e) &=& -k\,e\left({1\over 4}T^{abc}T_{abc} + {1\over 2}T^{abc}T_{bac} 
- T^{a}T_{a}\right) \nonumber \\
& \equiv & -ke\Sigma^{abc}T_{abc} \,,
\label{3}
\end{eqnarray}
where $k = c^3/(16\pi G)$,  and $\Sigma^{abc}$ is defined by

\begin{equation}
\Sigma^{abc} = {1\over 4}\left(T^{abc} + T^{bac} - T^{cab}\right) 
+ {1\over 2}\left(\eta^{ac}T^{b} - \eta^{ab}T^{c}\right)\,.
\label{4}
\end{equation}
In the expressions above, we have the following definitions:
$ T_{a} = T^{b}\,_{ba}$ and 
$T_{abc} = e_{b}\,^{\mu}e_{c}\,^{\nu}T_{a\mu\nu}$. 

The field equations obtained after the variation of Eq. (\ref{3}) with 
respect to $e^{a\mu}$ are

\begin{equation}
e_{a\lambda}e_{b\mu}\partial_\nu (e\Sigma^{b\lambda \nu} )-
e (\Sigma^{b\nu}\,_aT_{b\nu\mu}-
{1\over 4}e_{a\mu}T_{bcd}\Sigma^{bcd} )=0\,.
\label{5}
\end{equation}
The field equations above are equivalent to Einstein's equations for
the metric tensor, are invariant under local Lorentz transformations, 
and can be rewritten in a remarkable simple form,

\begin{equation}
\partial_{\nu}(e\Sigma^{a\mu\nu}) = 
{1\over 4k}ee^{a}\,_{\nu}(t^{\mu\nu} )\,,
\label{6}
\end{equation}
where 
\begin{equation}
t^{\mu\nu} = k(4\Sigma^{bc\mu}T_{bc}\,^{\nu} - 
g^{\mu\nu}\Sigma^{bcd}T_{bcd})\,,
\label{7}
\end{equation}
which is interpreted as the gravitational energy-momentum tensor. Had we
considered matter fields in the expressions above, we would just make the
substitution 
$t^{\mu\nu}\rightarrow t^{\mu\nu}+{1\over c}\texttt{T}^{\mu\nu}$ in Eq.
(\ref{6}), the latter
quantity being the energy-momentum tensor for the matter fields. Again, we
refer to the reviews \cite{TEGR-1} and the recent one \cite{TEGR-2} for 
details and further explanations. A full discussion of all consistency
properties regarding the field equations of the TEGR is given in the
review article \cite{TEGR-2}.

The field equations (\ref{6}) yield the total energy-momentum of
the gravitational field, which is defined primarily 
as a surface integral whose 
boundary $S$ encompasses an arbitrary volume $V$ of the three-dimensional
space. We establish the definition

\begin{equation}
P^a=\int_V d^3x\,e\,e^a\,_\mu t^{0\mu}= -\oint_S dS_j\,\Pi^{aj}\,,
\label{8}
\end{equation}
where $\Pi^{aj}=-4ke\,\Sigma^{a0j}$. In the Hamiltonian formulation of the
theory, $\Pi^{aj}$ is the momentum canonically conjugated to $e_{aj}$.
This definition may be easily applied to physical situations provided we 
choose a timelike congruence 
of vector fields $e_{(0)}\,^\mu$ adapted to suitable 
observers in space-time, i.e., (i) static observers, (ii) observers in 
stationary state, (iii) in free fall, (iv) dragged by the rotational motion
of the sources, etc. In similarity to the situations in classical
and/or relativistic theories, the notion of energy in the TEGR depends on 
the frame of an observer in space-time. This definition is consistent
and has been satisfactorily verified in several applications \cite{TEGR-2},
including the Bondi-Sachs space-time.

\subsection{The Bondi-Sachs space-time and its energy-momentum}

The Bondi-Sachs space-time \cite{Bondi,Sachs} is important, both from the 
historical and physical points of view, because it is a self-consistent 
formulation that shows that gravitational radiation exists in space-time due
to a loss of mass of the source. To our knowledge, no better formulation has
emerged in the literature that explains such an important phenomenon in the
universe. The  Bondi-Sachs line element is ordinarily presented in the 
($u$, $r$, $\theta$, $\phi$) coordinates, where $r$ is the ordinary radial 
variable and $u$ is the retarded time, $u=t-r$ ($r$, $\theta$ and $\phi$
are spherical coordinates), and is constructed in terms of three functions,
$M(u,\theta,\phi)$, which is the {\it mass aspect}, and $c(u,\theta,\phi)$, 
$d(u,\theta,\phi)$, which are related to the news functions.

The Bondi-Sachs line element is not given in closed mathematical form. It is
usually displayed in powers of the radial coordinate $r$. There are very good
review articles that address the Bondi-Sachs line element as well as the
Bondi-Sachs energy-momentum vector $m_\mu(u)$
\cite{Trautman,Pirani,Sachs-2,Goldberg}. The latter quantity is usually 
defined only in terms of the integral of the mass aspect 
$M(u,\theta,\phi)$, and not in terms
of the functions $c(u,\theta,\phi)$ and $d(u,\theta,\phi)$ (see, for instance, 
Eq. (4.4) of Ref. \cite{Goldberg}). This is a surprising feature, because the
news functions $\partial_0 c$ and $\partial_0 d$ are expected to yield
physical manifestations of the gravitational field, both at spacelike or null
infinities. In Ref. \cite{Maluf2} we have readdressed this issue
in the context of the TEGR, and found 
that in addition to the integral of the mass aspect, there is a contribution
to the gravitational energy-momentum of the Bondi-Sachs space-time that is 
constructed exclusively in terms of the functions $c$ and $d$. We call this
extra term as the energy of gravitational radiation, because
it is totally detached from the energy of the source. In the definition of
the energy of gravitational radiation there is no dependence in the radial
coordinate $r$. It is simply a quantity spread all over the space-time, 
although an integration in the angular coordinate $\theta$ is made (we will
restrict considerations to axial symmetry, as we will explain).

We will follow the presentation displayed both in Refs. \cite{TEGR-2} and
\cite{Maluf2}. In the standard presentation, the Bondi-Sachs line element is 
given by

\begin{eqnarray}
ds^2&=&g_{00}\,du^2+g_{22}\,d\theta^2+g_{33}\,d\phi^2 \nonumber \\
&{}&+2g_{01}\,du\,dr+2g_{02}\,du\,d\theta+2g_{03}\,du\,d\phi+
2g_{23}d\theta\,d\phi\,, 
\label{9}
\end{eqnarray}
where

\begin{eqnarray}
g_{00}&=&{V\over r}e^{2\beta}-r^2(e^{2\gamma}U^2\cosh 2\delta 
+e^{-2\gamma}W^2\cosh 2\delta +2UW \sinh 2\delta)\,, \nonumber \\
g_{01}&=& -e^{2\beta}\,, \nonumber \\
g_{02}&=&-r^2(e^{2\gamma}U\cosh 2\delta+W\sinh 2\delta)\,, \nonumber \\
g_{03}&=&-r^2\sin\theta(e^{-2\gamma}W\cosh 2\delta 
+U \sinh 2\delta)\,, \nonumber \\
g_{22}&=&r^2e^{2\gamma}\cosh 2\delta\,, \nonumber \\
g_{33}&=&r^2e^{-2\gamma}\cosh 2\delta\,\sin^2 \theta\,, \nonumber \\
g_{23}&=&r^2\sinh 2\delta\,\sin\theta\,.
\label{10}
\end{eqnarray}
We adopt the usual convention $(u,r,\theta,\phi)=(x^0,x^1,x^2,x^3)$.
The functions $\beta$, $\gamma$, $\delta$, $U$ and $W$ in the equations above
are not exact. They are given only in asymptotic form, in powers of $1/r$. 
We will dispense with the powers of $1/r$ of 
the field quantities that do not contribute to the calculations. Thus, the 
asymptotic form of the functions above are

\begin{eqnarray}
V&\simeq& -r+2M\,, \nonumber \\
\beta &\simeq& -{{c^2+d^2}\over {4r^2}}\,,\nonumber \\
\gamma &\simeq& {c\over r}\,, \nonumber \\
\delta &\simeq& {d\over r}\,, \nonumber \\
U&\simeq& -{{l(u,\theta,\phi)}\over r^2}\,, \nonumber \\
W&\simeq& -{{\bar{l}(u,\theta,\phi)} \over r^2}\,,
\label{11}
\end{eqnarray}
where 
$$l=\partial_2 c+2c\,\cot\theta+\partial_3 d\,\csc \theta\,,$$
$$\bar{l}=\partial_2 d +2d\,\cot\theta -\partial_3 c\csc \theta\,.$$
In the limit $r\rightarrow \infty$, the asymptotic form of the functions above
yield

\begin{eqnarray}
g_{00}&\simeq & -1+{{2M}\over r}\,, \nonumber \\
g_{01}&\simeq & -1+{{c^2+d^2}\over {2r^2}}\,,\nonumber \\
g_{02}&\simeq & l+{1\over r}(2cl +2d\bar{l} -p)\,, \nonumber \\
g_{03}&\simeq & \bar{l}\sin\theta + 
{1\over r}(-2c\bar{l}+2dl -\bar{p})\sin\theta\,, \nonumber \\
g_{22}&\simeq & r^2+ 2cr+2(c^2+d^2)\,, \nonumber \\
g_{33}&\simeq & \lbrack r^2- 2cr+2(c^2+d^2)\rbrack\sin^2\theta\,,\nonumber \\
g_{23}&\simeq & 2dr\sin\theta+{{4d^3}\over {3r}} \sin\theta\,.
\label{12}
\end{eqnarray}
The functions $p$ and $\bar{p}$ are defined in Refs. \cite{Z1,Z2}. They 
depend on functions that are not defined above. However, they will not 
contribute to the final expressions, and for this reason we will not present 
their definitions here.

As explained in Refs. \cite{Maluf2,TEGR-2}, definition (\ref{8}) is applied
to the Bondi-Sachs line element in the context of a field of static observers
determined by the condition $e_{(0)}\,^i=0$. Although conceptually this is a
very simple procedure, it requires a careful handling of the long algebraic
expressions. If no symmetries are assumed in the space-time, the
final expression for the gravitational energy in the 
Bondi-Sachs space-time is \cite{Maluf2,TEGR-2}

\begin{equation}
P^{(0)}=4k\int_0^{2\pi} d\phi \int_0^{\pi}d\theta \sin\theta
\biggl[M+\partial_0 F\biggr]\,, 
\label{14}
\end{equation}
where

\begin{equation}
F=-{1\over 4}\biggl(l^2 + \bar{l}^2 \biggr) +{1\over 2}c^2 +d^2\,.
\label{15}
\end{equation}
The quantities under integration in the expression above are the only ones
that contribute to the surface integral in the limit $r\rightarrow \infty$.

The first term in Eq. (\ref{14}) is the ordinary expression for the 
gravitational energy of the Bondi-Sachs space-time, as displayed in the 
standard literature on the subject. It is given by the integration of the 
mass aspect $M(u,\theta,\phi)$ only.

The new term $\partial_0 F$ generalises the ordinary Bondi-Sachs energy.
This term does not depend on the mass aspect, but only on the functions 
$c(u,\theta,\phi)$ and $d(u,\theta,\phi)$. Assuming the speed of light $c=1$
as well as $G=1$, the quantity \cite{Maluf2,TEGR-2},

\begin{equation}
E_{rad}={1\over {4\pi}}\int_0^{2\pi} d\phi \int_0^{\pi}d\theta \sin\theta\,
(\partial_0 F)\,,
\label{16}
\end{equation}
is interpreted as the energy of the gravitational radiation, i.e., a form
of energy detached from the source.

Before closing this section, we recall the important relationship between the
time derivative of the mass aspect with the functions $c$ and $d$, obtained 
directly from the field equations.  This relationship is given by 
\cite{Sachs,Sachs-2}

\begin{equation}
\partial_0 M=-\lbrack (\partial_0c)^2+(\partial_0 d)^2\rbrack
+{1\over 2}\partial_0\biggl(\partial_2 l+l\cot\theta+
{{\partial_3 \bar{l}}\over {\sin\theta}}\biggr)\,.
\label{17}
\end{equation}
We see that the existence of the news functions $\partial_0 c$ and 
$\partial_0 d$ imply a variation of the mass aspect, and
consequently leads to a variation of the space-time geometry. This variation
may occur in the form of propagating ripples in the space-time, which is
precisely the nature of radiation.

\section{Energy of the gravitational radiation and noise in the gravitational wave detectors}

In this section we will obtain an explicit expression and values
for the energy of 
gravitational radiation, by imposing axial symmetry to the gravitational
field, i.e., we will require $d=0=\bar{l}$. Thus, the field quantities do not
depend on the variable $\phi$. We will further adopt a form of the news
function first considered by Papapetrou \cite{Papapetrou}, Hallidy and Janis
\cite{Janis}, and Hobill \cite{Hobill}. By still adopting the usual 
convention $x\equiv \cos\theta$, we will consider the news function $c(u,x)$
to be given as

\begin{equation}
c(u,x)=(1-x^2)\sum_{n=2}a_n(u){{d^2}\over {d x^2}}P_n(x)\,,
\label{18}
\end{equation}
where $P_n(x)$ are the Legendre polynomials and $a_n(u)$ are general 
functions of the retarded time $u$, that reduce to constants when the source
is not radiating. Of course, it is assumed that the sum  in the 
expression above converges. Definitely, it is not trivial to choose a 
suitable form of the news function. We will rely on the work developed by
the authors mentioned above. As far as the expression of the mass aspect
$M(u,\theta)$ is not a priori specified, Eq. (\ref{18}) is perfectly valid.

We first obtain an expression for the quantity $F$ considering axial 
symmetry, i.e., $d=0=\bar{l}$.
It reads \cite{Maluf2}

\begin{equation}
F=-\frac{(\partial_\theta c)^2}{4}-\cot\theta\, c(\partial_\theta c)
-\frac{3}{2}\,c^2\cot^2\theta+\frac{c^2}{2\sin^2\theta}\,.
\label{19}
\end{equation}
By making use of some identities between the Legendre polynomials, we 
rewrite the news function as

\begin{equation}
c(u,x)=\sum_{n=2} a_n(u)
\biggl[2x{d\over {{dx}}}P_n(x)-n(n+1)P_n(x)\biggr]\,.
\label{20} 
\end{equation}
Substitution of Eqs. (\ref{19}) and (\ref{20}) into Eq. (\ref{16}) 
yields \cite{Maluf2}

\begin{equation}
E_{rad}=\frac{1}{8}\partial_0\int_{-1}^{1}\left[-(1-x^2)(\partial_x c)^2+
4x\,c\,\partial_x c+2c^2(1-3x^2)\right]dx\,.
\label{21}
\end{equation}

The final form of the energy of gravitational radiation is obtained after
two more steps of simplifications, as shown in Ref. \cite{Maluf2} but omitted
here. The final form is given by

\begin{eqnarray}
E_{rad}&=&-\frac{1}{8}\partial_0\sum_{n=2}\frac{a_n^2(n+2)!
(16n^4+28n^3+20n^2+11n-105)}{(n-2)!(2n+1)(2n-1)(2n+3)}\nonumber \\
&&+\frac{1}{8}\partial_0\sum_{n,m=2}\frac{a_na_m(n+2)![1+(-1)^{n+m}]}{(n-2)!}
\nonumber\\
&&-\frac{3}{2}\partial_0\sum_{n=2}\frac{a_na_{n+2}(n+4)!}
{(n-2)!(2n+1)(2n+3)(2n+5)}\,.
\label{22}
\end{eqnarray}

In what follows, we will choose suitable forms for the functions $a_n(u)$ and
obtain graphical expressions for the energy $E_{rad}$. In the three cases
considered below, there appears a term $1/n!$ in the expressions of $a_n(u)$
that guarantees the convergence of the series in Eq. (\ref{22}). 
Quantities other than $n!$ would also guarantee the convergence, but we 
choose the latter quantity just for simplicity. The three cases to be 
considered are the following:

\begin{eqnarray}
\texttt{Case 1}&:& \ \ \
a_n(u)={1\over{n!}} \biggl[{1\over \omega} e^{-u^2\omega^2} \biggr] \ 
\equiv {1\over {n!}}\, f_1(u,\omega) \,, 
\label{23} \\
\texttt{Case 2}&:& \ \ \
a_n(u)= {1\over {n!}} \biggl[{1\over \omega} 
e^{-u^2/ \lambda^2} \sin{\omega u}
\biggr] \  \equiv {1\over {n!}} f_2(u,\omega)\,, 
\label{24} \\
\texttt{Case 3}&:& \ \ \
a_n(u)= {1\over {n!}} \biggl[{1\over \omega} \sin^2{\omega u}\biggr] \ \equiv
{1\over {n!}} f_3(u,\omega)\,.
\label{25}
\end{eqnarray}
The equations above define the functions 
$f_1(u,\omega), f_2(u,\omega), f_3(u,\omega)$.
The quantity $\omega$ is an angular frequency specified in each of the 
figures below, and whose value is one order of magnitude higher than the
frequency of the expected linearised gravitational wave that are measured 
in the large terrestrial laboratories LIGO-Virgo-KAGRA. The attenuation 
parameter $\lambda$ in eq. (\ref{24}) is independent of $\omega$, and takes
the value $\lambda=0.06$ in natural units.

Let us first consider Case 1.
The energy of the gravitational radiation may be expressed by

\begin{equation}
\label{26}
E_{rad} = \partial_0 (f_1(u,\omega))^2 (A+B+C)\,,
\end{equation}
where 

\begin{eqnarray}
\label{27}
A&=&-\frac{1}{8}\sum_{n=2}\left[\frac{(n+2)!(16n^{4}+28n^{3}+
20n^{2}+11n-105)}{[(n!)^2(n-2)!(2n+1)(2n-1)(2n+3)]}\right]\,, \\
\label{28}
B&=& \frac{1}{8}\sum_{n,m=2}
\left[\frac{(n+2)![1+(-1)^{n+m}]}{n!m!(n-2)!}\right]\,, \\
\label{29}
C&=& -\frac{3}{2}\sum_{n=2}\left[\frac{(n+4)!}
{n!(n+2)!(n-2)!(2n+1)(2n+3))(2n+5)}\right]\,,
\end{eqnarray}
and $f_1(u,\omega)$ is defined in Eq. (\ref{23}). By means of numerical
(computer) evaluation, we find

\begin{eqnarray}
A&=& -7.04\,, \label{30} \\
B&=& 5.21\,, \label{31} \\
C&=& -0.088\,. \label{32} 
\end{eqnarray}
Thus, Eq. (\ref{26}) is rewritten as 

\begin{equation}
\label{33}
E_{rad} = -1.92\, \partial_0 (f_1(u,\omega))^2\,.
\end{equation}
Similarly, for the Cases 2 and 3 we have, respectively,

\begin{equation}
\label{34}
E_{rad} = -1.92 \, \partial_0 (f_2(u,\omega))^2\,,
\end{equation}

\begin{equation}
\label{35}
E_{rad} = -1.92 \, \partial_0 (f_3(u,\omega))^2\,.
\end{equation}
The values of gravitational energy above are given in natural units. In order
to transform to SI units, we just make the substitution

$$E_{rad} \rightarrow E_{rad} \,{c^4\over G}\,,$$
where $c$ is the speed of light and $G$ is Newton's constant. The graphical
expressions of $E_{rad}$ as a function of the retarded time $u=t-r$ are given
below, in Figures 1, 2, 3, which are presented in natural units.

\begin{figure}[h]
\centering
\includegraphics[width=0.7\textwidth]{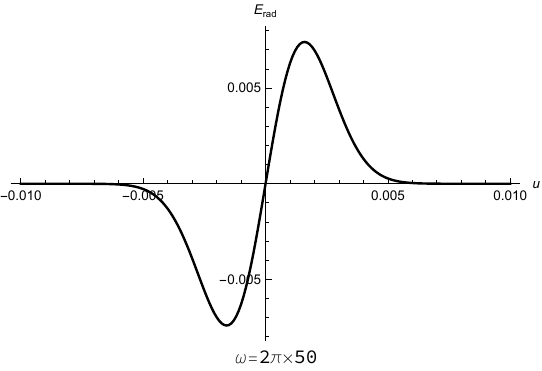}
\caption{Energy of the gravitational radiation; $a_n(u)$ is given in Case 1}
\end{figure}

\begin{figure}[h]
\centering
\includegraphics[width=0.7\textwidth]{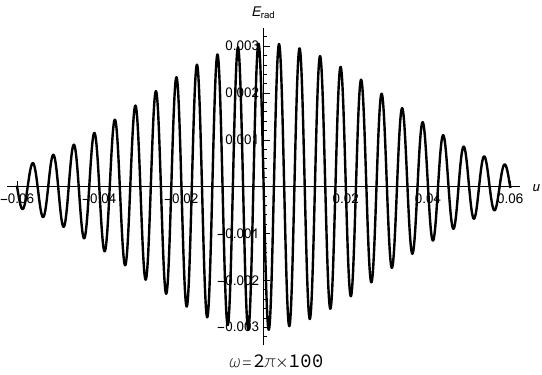}
\caption{Energy of the gravitational radiation; $a_n(u)$ is given in Case 2
and $\lambda=0.06$ in natural units.}
\end{figure}

\begin{figure}[h]
\centering
\includegraphics[width=0.6\textwidth]{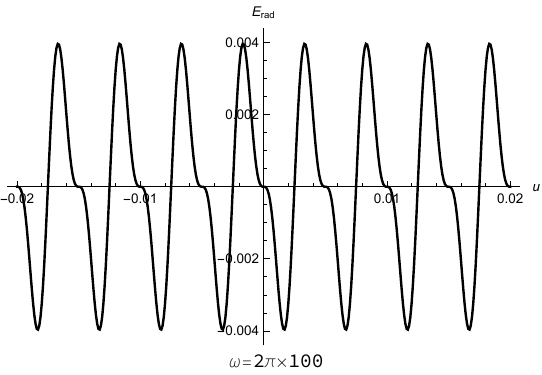}
\caption{Energy of the gravitational radiation; $a_n(u)$ is given in Case 3.}
\end{figure}

In Case 1, the time dependent part of the news function is chosen 
as a simple Gaussian. In Case 2, it is given by a Gaussian multiplied by a 
harmonic function, and in Case 3, the time dependent part of 
$a_n(u)$ is given by a harmonic function only.
We believe that the Case 2 (Figure 2) is more realistic, i.e., it is close 
to depict an actual burst of gravitational radiation. 
But Case 3 (Figure 3) could as well 
represent a certain range that follows a violent astrophysical event that 
generates gravitational radiation. Due to its non-oscillatory behaviour, we
believe that Case 1 (Figure 1) does not to play a major role in the 
present analysis. Cases 2 and 3
represent sharp oscillations of the whole space-time energy as given 
by Eq. (\ref{14}). These oscillations travel through the entire space-time.
In regions very far from the source that originates the first term in 
Eq. (\ref{14}), they represent the background gravitational energy of an 
otherwise ideally flat space-time. Given that a huge, undefined number of 
astrophysical events generated and still generate
such a form of gravitational radiation
throughout the universe, the gravitational energy discussed above in
qualitative models should be travelling in space-time, in similarity to the
propagation of light rays from very distant stars that are visible on Earth,
and that may not be shining (and not existing) any longer.

The question arises as to whether this form of radiation can be measured. We
argue that this radiation might already have been detected in the large
terrestrial laboratories, LIGO-Virgo-KAGRA, in the form of noise. As we
discussed at the beginning of the article, it is presently assumed that the
exact nature of noise in these detectors is unknown. These detectors do not
measure directly gravitational energies such as discussed above, but they
are designed to measure gravitational waves expressed directly in the metric
tensor. Thus, the $g_{22}$ and $g_{33}$ components of the metric tensor
displayed in Eqs. (\ref{10}) and (\ref{12}) should manifest in the
LIGO-Virgo-KAGRA detectors. The relevant part of these metric components is
precisely the function $c(u,\theta)$. In the following, we will display some
figures of this function, starting with Case 1.

\begin{figure}[h]
\centering
\includegraphics[width=0.6\textwidth]{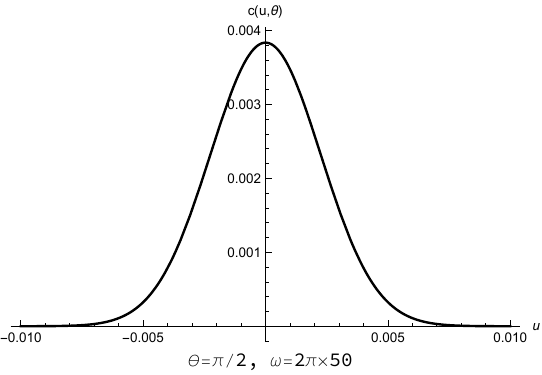}
\caption{News function, Case 1: $\theta=\pi/2$.}
\end{figure}

\begin{figure}[h]
\centering
\includegraphics[width=0.6\textwidth]{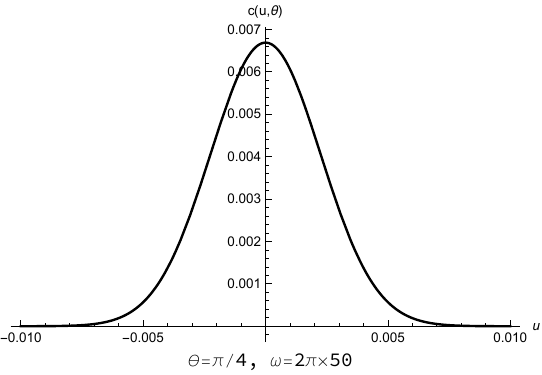}
\caption{News function, Case 1: $\theta=\pi/4$.}
\end{figure}

\begin{figure}[h]
\centering
\includegraphics[width=0.7\textwidth]{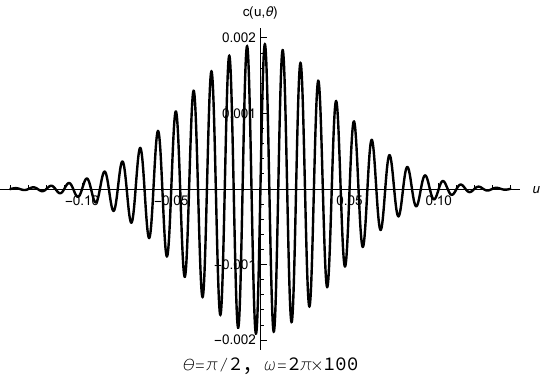}
\caption{News function, Case 2: $\theta=\pi/2$.}
\end{figure}

\begin{figure}[h]
\centering
\includegraphics[width=0.7\textwidth]{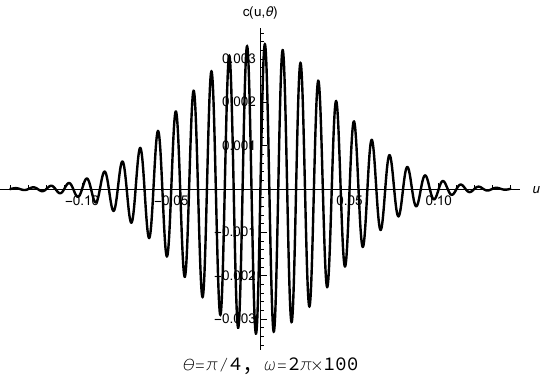}
\caption{News function, Case 2: $\theta=\pi/4$.}
\end{figure}

\begin{figure}[h]
\centering
\includegraphics[width=0.6\textwidth]{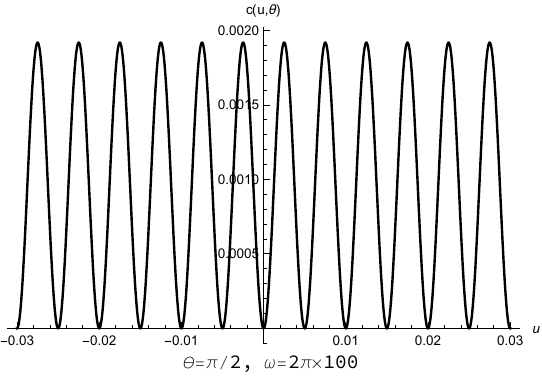}
\caption{News function, Case 3: $\theta=\pi/2$.}
\end{figure}

The figures relative to Case 1 (Figures 4, 5 and 9)
confirm that such a form of news function
is likely not contribute to the noise in the detectors, and therefore they
are not a description of a burst of gravitational radiation. Some sort of
non-trivial oscillation of the source should take place in the astrophysical
process, in similarity to electromagnetic radiation, which exists because of
a permanent oscillatory
(in general, harmonic) behaviour of the source. But Case 2
(Figures 6, 7, 10 and 11) and Case 3 (Figures 8 and 12) generate
physical effects that could already have been detected in the large
terrestrial laboratories in the form of noise. Of course, noise is not
the result of a single burst of gravitational radiation, but is a 
consequence of uncountable astrophysical processes generated in the entire
history of the universe. It may represent a background excitation of the 
whole space-time, even from the classical point of view, and is everywhere
present. 

Figures 9, 10, 11 and 12 display the angular dependence of the 
news function $c(u,\theta)$ in the interval $-1< x< +1$, where 
$x=\cos \theta$. These figures are not symmetrical around $x=0$, or
$\cos\theta=\pi/2$. But in all cases, the news function vanishes when 
$x=\pm 1$, in agreement with Eq. (\ref{18}).

\begin{figure}[h]
\centering
\includegraphics[width=0.7\textwidth]{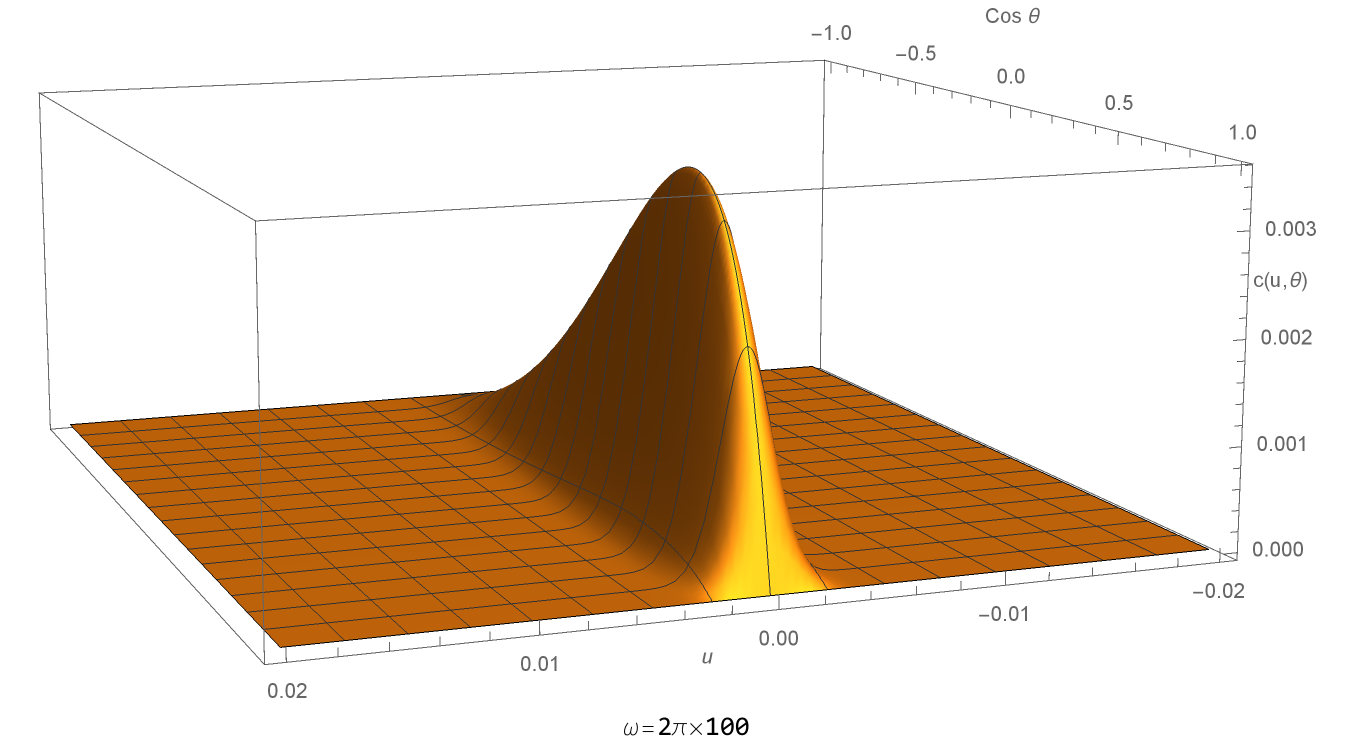}
\caption{News function, Case 1}
\end{figure}

\begin{figure}[h]
\centering
\includegraphics[width=1.0\textwidth]{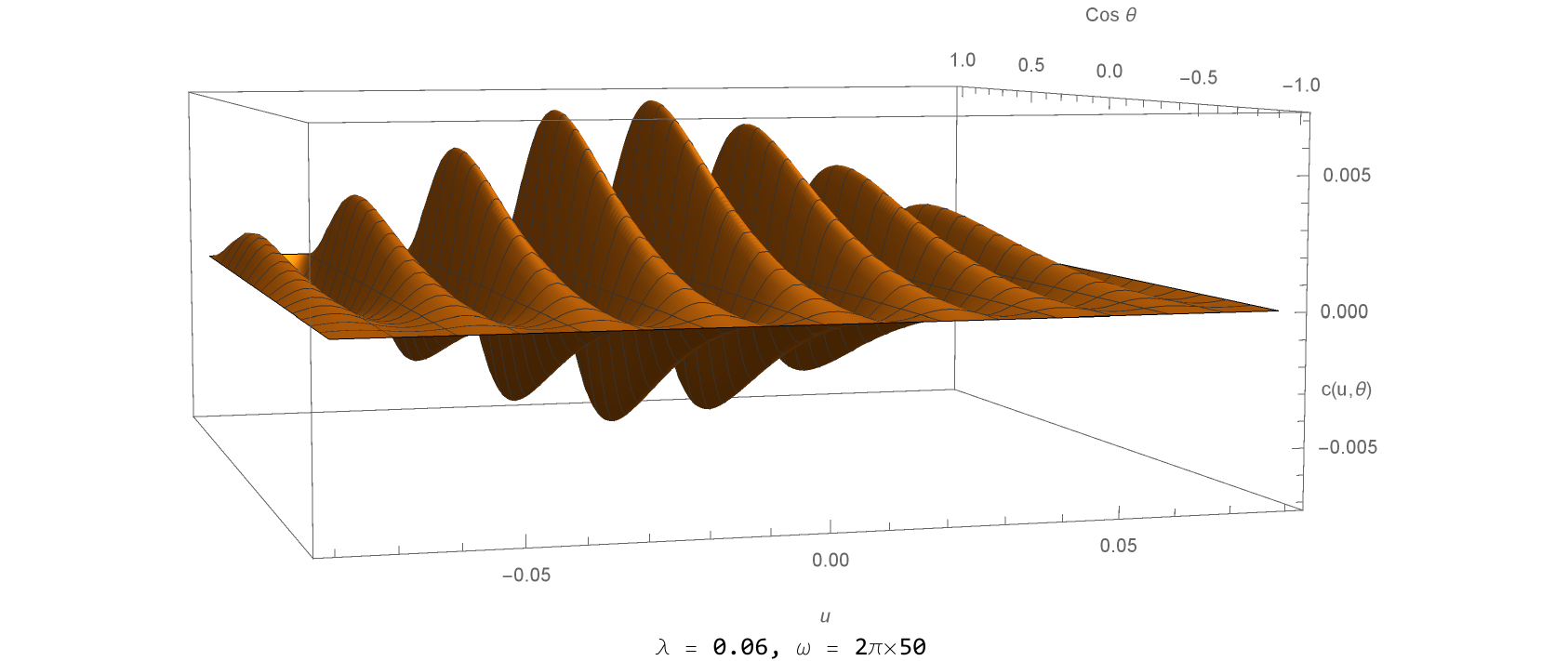}
\caption{News function, Case 2}
\end{figure}

\begin{figure}[h]
\centering
\includegraphics[width=1.0\textwidth]{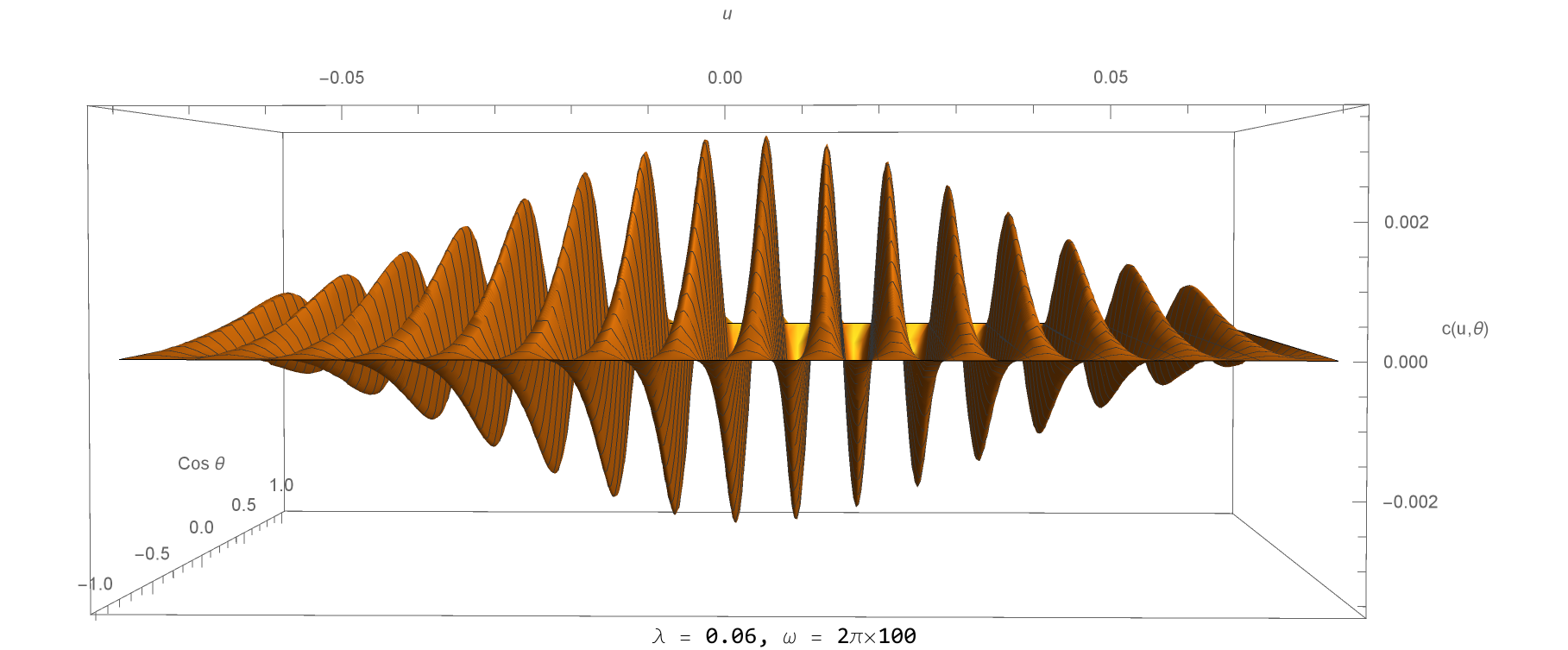}
\caption{News function, Case 2}
\end{figure}

\begin{figure}[h]
\centering
\includegraphics[width=0.8\textwidth]{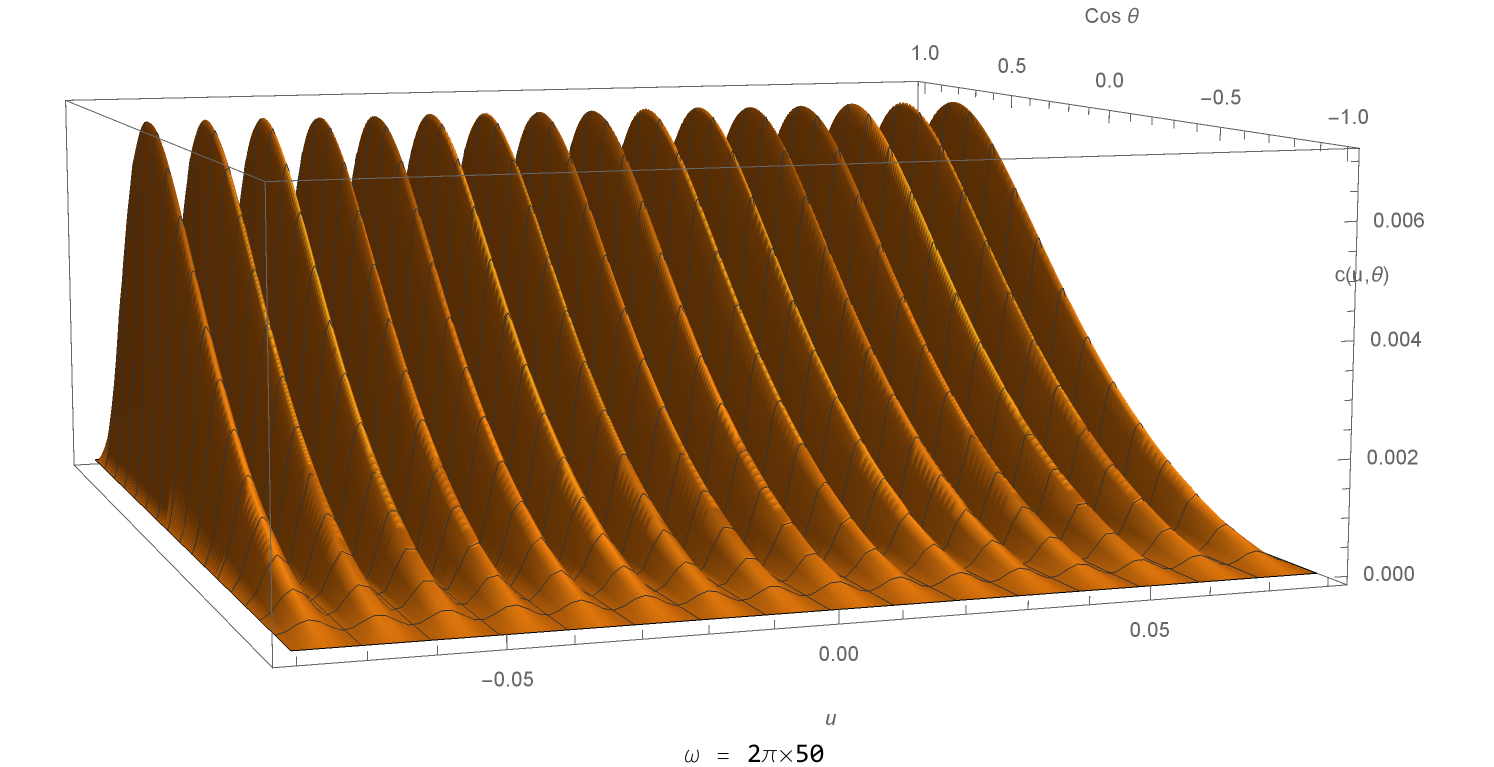}
\caption{News function, Case 3}
\end{figure}

The non-convergence of Eq. (\ref{18}) yields several difficulties, assuming
that the non-convergence yields an infinite value for the news function.
The first difficulty is that all observable quantities are finite, and the
non-convergence would imply an infinite value for $c(u,x)$, and consequently
a non-observable quantity. The second difficulty is connected to Eq. 
(\ref{17}). The latter equation relates the loss of mass with the news 
functions. Both quantities, the variation of loss of mass and the news 
functions are 
intertwined. So, the establishment of alternative news functions is not so 
straightforward, as it appears to be. The loss of mass is dictated by actual
astrophysical processes. To some extent, this issue was already considered 
by Papapetrou \cite{Papapetrou}, and Hallidy and Janis \cite{Janis}. In 
principle, we measure directly the news function, and then we infer the 
mathematical description of the loss of mass. Here we note that an infinite
value for the news function implies, via Eq. (\ref{17}), an ultra fast loss 
of mass, since $\partial_0 M \rightarrow -\infty$,
and consequently the immediate disappearance of the astrophysical 
source (a star, a binary system, or a more complex configuration). This ultra
fast loss of mass probably does not take place in nature, and therefore we
exclude the possibility of non-convergence of Eq. (\ref{18}). 

\section{Concluding remarks}

The present analysis has two major motivations. The first is the existence of
an expression for the energy of gravitational radiation that is predicted in 
the TEGR, and which depends only on the news functions $c(u,\theta,\phi)$ and
$d(u,\theta,\phi)$. This energy is not obtained in the context of the metric
formulation of general relativity. The standard Bondi-Sachs energy-momentum
is defined only in terms of the mass aspect $M(u,\theta,\phi)$, and such a
definition seems to be incomplete since the news functions are quantities that
are supposed to manifest at null or spacelike infinities, and 
that carry information about the source of gravitational radiation. 

The second motivation is the inconclusive work by Papapetrou 
\cite{Papapetrou} and Hallidy and Janis \cite{Janis}, who tried to prove (but
failed) that a stationary space-time may be the final stage of a space-time
in which a burst of gravitational radiation has taken place. Indeed, assuming
that gravitational radiation travels in the space-time similarly to 
electromagnetic radiation, then the space-time may become flat again if all
the radiation is dissipated by some unknown process. Otherwise, a background
radiation and background gravitational energy exist and could be detected.
This is precisely the argument presented in Section 3 of this article. 
One manifestation of the gravitational radiation, which is a form of 
gravitational field, is a substantial contribution to the noise that has
already been detected in the LIGO-Virgo-KAGRA detectors. These detectors are
designed to measure gravitational fields, and gravitational radiation is a 
form of gravitational field. If the gravitational radiation is not 
dissipated, then the idea of a perfectly flat space-time is just an 
idealization, albeit an important concept, but a perfectly flat space-time
might not exist at all. We cannot estimate the amount of
astrophysical explosions that have taken place in the universe in its whole
history. But it is really important to emphasize that a loss of mass of a
certain source necessarily implies in a change of geometry, implies in a
loss of energy of the source, and some energy must be released in the
form of gravitational radiation. This is the essence of the Bondi-Sachs
framework.

In Section 3, we have worked under the simplified assumption of axial 
symmetry. One realistic model for the news function $c(u,\theta)$ and for the
emitted gravitational radiation is given by Case 2. As we argued, Case 3 
could represent a realistic model for radiation, for a limited interval of
the radiation process. Radiation is always linked to the oscillations of the
source, exactly like in the generation of electromagnetic radiation.
In the two cases discussed in Section 3, Cases 2 and 3, 
we obtain an oscillating 
background gravitational energy. The oscillations are supposed to be very
tiny, and rather imperceptible, but would imply a lower, minimum value
for the temperature in the universe. The role of news functions of the type
displayed by Case 1 should be further analysed.
The investigations along the lines
described in this article could proceed by working out refinements in the
definition of the news function. The Bondi-Sachs framework could provide
a contribution to the ``missing physics" suggested in Ref. \cite{Gupta}.

\end{document}